\documentclass[aps,10pt,showpacs,twocolumn]{revtex4}
\usepackage{amsmath, amssymb, lipsum}
\usepackage{graphics, graphicx, epsfig, color, subfigure}
\begin{document}

\title{Density of Saturated Nuclear Matter at Large $N_{c}$ and Heavy Quark Mass Limits}

\date{\today} 

\author{Prabal Adhikari}
\email{prabal@umd.edu}
\affiliation{Maryland Center for Fundamental Physics and the Department of Physics, 
University of Maryland, College Park, MD 20742-4111}

\author{Thomas D. Cohen}
\email{cohen@umd.edu}
\affiliation{Maryland Center for Fundamental Physics and the Department of Physics, 
University of Maryland, College Park, MD 20742-4111}

\author{Ishaun Datta}
%\email{...}
\affiliation{Montgomery Blair High School, 51 University Boulevard East, Silver Spring, MD 20901-2451}

\begin{abstract}
We exhibit the existence of stable, saturated nuclear matter in the large $N_{c}$ and heavy quark mass limits of QCD. In this limit, baryons (with the same spin flavor structure) interact at leading order in $N_{c}$ via a repulsive interaction due to the Pauli exclusion principle and at subleading order in $1/N_c$ via the exchange of glueballs. Assuming that the lightest glueball is a scalar, which implies that the subleading baryon interaction  is attractive, we find that nuclear matter saturates since the subleading attractive interaction is longer ranged than the leading order repulsive one.  We find that the saturated matter is in the form of a crystal with either a face-centered cubic or a hexagonal-close-packed symmetry with  baryon densities of $\mathcal{O}((\, \tilde{\alpha}_{s} m_q (\ln (N_{c}m_{q}\Lambda_{\textrm{QCD}}^{-1}))^{-1})^3 )$. Remarkably, the leading order expression for the density of saturated nuclear matter is independent of the lighest glueball mass and scalar-glueball-baryon coupling in the extreme large $N_{c}$ limit or heavy quark limit (or both), which we define precisely in this work.
\end{abstract}
\maketitle
%\section{Introduction}

QCD has been known to be the theory of strong interactions for over forty years. However, it is still not possible to study saturated nuclear matter directly from the QCD lagrangian. In principle, the lagrangian encodes information about all the possible phases of QCD in a variety of physical environments~\cite{stephanov}, many of which are known already with a reasonable amount of certainty. However, there are significant barriers in understanding even the most basic properties of nuclear matter such as saturation.

Contemporary methods for studying QCD are not adequate for the problem of saturated nuclear matter. Lattice QCD is the only viable known method for investigating properties of QCD from first principles~\cite{latticereview}. It has been very successful in the prediction of masses of stable, low-lying hadrons, e.g. pions and the baryon octet and low-energy phase shifts in hadron scattering. With ever increasing computational capabilities, one expects lattice QCD to determine masses and other physical observables with increasing accuracy. However, saturated nuclear matter cannot be studied on the lattice regardless of computational capacity. The Monte-Carlo methods associated with lattice QCD require the fermion determinant to be positive-definite; unfortunately, the fermion sign problem renders these probabilistic methods completely useless~\cite{signproblem1, signproblem2}.
 
An alternative method for studying QCD involves the use of effective field theories (EFTs), which can generally be used for problems involving large scale separations. This approach captures symmetries of the original theory in the relevant regime where the EFT is applicable. For example, chiral perturbation theory is a systematic method for studying the behavior of the lightest hadrons i.e. the pions, which are the Goldstone bosons associated with the breaking of chiral symmetry. EFTs are appealing in that physical phenomena can be studied in a systematic manner with errors that are controlled and physically well-understood. Historically, nuclear matter was studied in the context of nuclear models but they were often unsystematic and had uncontrolled errors. In order to characterize nuclear matter, one first needs to understand the many-body theory of nucleon interactions. However, the problem of nuclear matter is inherently non-perturbative, since nuclear matter forms a bound state; furthermore, scattering of nucleons is non-perturbative due to the possible formation of weakly bound nucleon states (e.g. deuteron). Weinberg~\cite{weinberg}, proposed a method, where one constructs the most general lagrangian that encapsulates all the symmetries of QCD at low energies, then generates nucleon-nucleon potentials in a perturbative fashion and proceeds to solve the Lippman-Schwinger equation, which is inherently non-perturbative. The method has been successful in understanding the nature of few-nucleon interactions; for instance, scattering phase shifts and mixing parameters~\cite{deuteron} have been used to fit low-energy constants that arise in the EFT and used to predict physical observables such as deuteron binding energies~\cite{deuteron}. However, the general problem of saturated nuclear matter involves an extremely large number of nucleons. This requires understanding of many-body forces well beyond the two-body and three-body interactions we understand currently.
\newline
\newline

In light of difficulties associated with studying nuclear matter in real QCD, we consider a setting where the problem is actually tractable, namely the combined 't Hooft large $N_{c}$~\cite{tHooft1, tHooft2, manohar} and Witten heavy quark mass limits~\cite{witten}.  While this limit of QCD clearly does not describe the physical world, it is attractive to study from a theoretical perspective because many features of the theory become tractable.  It is hoped that studies in this limit might give insight into the nature of saturated matter.

In this combined limit, baryons are completely antisymmetric states of $N_{c}$ non-relativistic quarks, with each quark moving in a mean-field potential generated by the remaining quarks~\cite{coherentstates}. Baryon masses are of $\mathcal{O}(N_{c}m_{q})$ and the baryon charge density is sharply peaked in this limit with a width of $\mathcal{O}\left(\frac{1}{\tilde{\alpha_{s}}m_{q}}\right)$~\cite{3+1,1+1}, where $m_{q}$ is the quark mass and $4\pi\tilde{\alpha_{s}}\equiv g^{2}N_{c}$; in the 't Hooft large $N_{c}$ limit, $4\pi\tilde{\alpha_{s}}$ is kept fixed at $\mathcal{O}(1)$ as $N_{c}\rightarrow\infty$ ensuring that the QCD beta function does not become divergent and remains well-defined. Then nuclear matter can be formed with the heavy quark, large $N_{c}$ baryons as constituents. At leading order in $N_{c}$, baryons (all with the same spin-flavor structure) form a phase of infinite nuclear matter under external pressure. This should not be surprising: baryons are fermions and therefore, in the presence of other baryons with the same spin-flavor structure, due to Pauli repulsion, will assume a spatial state that is maximally orthogonal to the neigboring baryons. Since the kinetic energies of baryons are suppressed both by the large $N_{c}$ and heavy quark mass limits, crystalization is inevitable. It is important to note that crystalization happens even away from the heavy quark mass limit. e.g. Skryme crystals~\cite{skyrme}. In the large $N_{c}$ limit, there is an attractive interaction between baryons associated with the lightest mesons i.e. pions; the interaction becomes parametrically large with decreasing separation and therefore saturation is inevitable.

Saturation does not occur in the combined heavy quark and large $N_{c}$ limits at leading order in $N_{c}$. Due to Pauli repulsion, baryons repel each other and move infinitely apart.  Nuclear matter in that limit requires an external pressure to stabilize it.  {\it A priori} it might seem that the inclusion of higher order corrections should not change this.  Typically one expects an expansion is valid only in regimes in which the inclusion of a higher-order term does not substantially change the properties of the ground state.  However, in the present context the inclusion of subleasing effects in $N_c$ causes an important qualitative difference: nuclear matter saturates.  Mathematically, this is possible because while the interactions can be written as Taylor expansion in $1/N_c$, the saturation density cannot: it turns out to be nonanalytic.  Physically, the reason for this is that at the first subleading order in $1/N_c$ baryons interact via the exchange of glueballs, which are the lightest particles; in the large $N_{c}$ QCD spectrum with physical masses, pions are the lightest particles.  However, since this system is also in the heavy quark mass limit, pion masses are of $\mathcal{O}(N_{c}^{0}m_{q})$ and glueballs are the lightest states in the spectrum.  As long as the lightest glueball is a scalar, the longest-range interaction between baryons is attractive and exponentially large relative to interactions via other channels at the longest distances.  Thus one has the longest-ranged interaction being subleading in $1/N_c$ (although it has a strength of order $N_c^0$).  Now formally, it is known that if one has a system of particles that have a long-range attractive interaction with a strength of order one in some counting scheme and some particles which are parametrically heavy, the system must be self bound as the mass of the particles goes to infinity.  Since the nucleon is heavy in our combined limit, an attractive force of strength of order unity (as is the case for glueball exchange) is sufficient to bind them, ensuring saturated matter.   The distance scale at which the subleading but longer-ranged attraction matches the shorter-ranged but stronger (i.e. leading order) repulsion determine the saturation density.

Note that in real QCD, glueball states are somewhat subtle; the only viable method to determine the properties quantitatively is using lattice QCD~\cite{glueballinlattice}.  Note, however, that in this context the question of whether they exist is poorly-posed: glueball states mix with mesons that share the same quantum numbers and in any event are unstable. However, in the heavy quark mass limit, meson masses are of $\mathcal{O}(m_{q})$ and are pushed out to infinity.  Therefore, in this limit, glueball states become well-defined for arbitrary number of colors.   Moreover, in the large $N_{c}$ limit~\cite{tHooft1,tHooft2}, glueballs are stable with decay widths of $\mathcal{O}\left(\frac{1}{N_{c}^{2}}\right)$. Furthermore, it has been shown in a model independent way (in pure Yang-Mills, where glueballs are stable) through the use of QCD inequalities that the lightest glueballs are parity positive scalars~\cite{west, nair, nussinov}. Strictly speaking, they could be degenerate with other glueballs; however, under the standard assumption that additional symmetries are required to give rise to such degeneracy, we ignore the possibility. Scalar glueballs remain the lightest even in the case of the 't Hooft large $N_{c}$ limit of QCD; further evidence for this comes from lattice calculation~\cite{dalley, teper} of glueball masses using a large number of colors (upto $N_{c}=8$).
\\
\\

Assuming that the scalar glueball is the lightest particle in the spectrum of large $N_{c}$ QCD, our goal here is to determine the density of saturated nuclear matter in the combined heavy quark and large $N_{c}$ limits.  In doing so we will first solve a toy problem in which the baryons are all in the same spin-flavor state.  This ensure that the repulsion due to the Pauli principle at the quark level affects all pairs of nucleons.  We will relax this assumption toward the end of the paper.

 Since the nucleons are heavy, the kinetic energy of the nucleons does not play a role.  Thus, the properties of saturation are determined by the interaction  energy.  Moreover, one expects the interaction to have a distance which maximizes the attraction and which becomes strongly repulsive if the distance is shortened by even a small amount.  Thus, it is sensible to imagine the nucleons forming a crystal with this distance as the nearest neighbor separation.  To visualize this
 consider a physical problem, where we put a large number of baryons, assumed to be $B$, together and proceed to determine the density of such baryons at saturation. Thus for the purposes of determining the average density of saturated nuclear matter, it is sensible to treat the baryons as though they were hard spheres with each baryon occupying a spherical volume with a radius, which is half the nearest neighbor distance.  The total volume occupied by the $B$ baryons at saturation is denoted $V$. Then the density of saturated nuclear matter is:
\begin{equation}
\rho_{\textrm{SNM}}=\frac{B}{V/B}=\frac{1}{\frac{4\pi}{3 P}\left(\frac{d_{\textrm{SNM}}}{2}\right)^{3}}\ ,
\end{equation}
where $d_{\textrm{SNM}}$ is the separation of nearest neighbor baryons at saturation and $P$ is the packing factor. This factor is determined by the type of crystalline structure saturated nuclear matter assumes. Before discussing what the crystalline structure of saturated nuclear matter should be, we first proceed to determine the separation, $d_{\textrm{SNM}}$,  by minimizing the energy per baryon.
\\
\\

First we consider the explicit form for energy of nuclear matter in the toy problem, $E_{\textrm{NM}}^{\textrm{toy}}$, at sub-leading order in $N_{c}$. Recall that in this toy problem all nucleons are constrained to be in the same spin-flavor state.  We consider the toy problem first as it is slightly simpler conceptually then the fully unconstrained problem.  We will turn to the full problem later.  We express the quantity $E_{\textrm{NM}}^{\textrm{toy}}$ in terms of the characteristic length scale in the problem, $\frac{1}{\tilde{\alpha}_{s}m_{q}}$, to define the following dimensionless parameter:
\begin{equation}
\label{tilded}
\tilde{d}=\tilde{\alpha}_{s}m_{q}d\ ,
\end{equation}
where $m_{q}$ is the quark mass (note that we are in the heavy quark mass limit, which means $\frac{m_{q}}{\Lambda_{\textrm{QCD}}}\gg 1$) and $\tilde{\alpha}_{s}=\frac{g^{2}N_{c}}{4\pi}$, is the strong coupling constant with the appropriate $N_c$ scaling to ensure that $\tilde{\alpha}_{s}$ is independent of $N_c$~\cite{3+1}.

The ground state energy (per baryon) in the toy problem consists of the following contributions:
\begin{equation}
\label{en}
\frac{E_{\textrm{NM}}^{\textrm{toy}}(\tilde{d})}{B}=\frac{E_{\textrm{Pauli}}(\tilde{d})}{B}+\frac{E_{\textrm{gb}}(\tilde{d})}{B}%+\frac{E_{D}}{B}.
\end{equation}
The first piece, $\frac{E_{\textrm{Pauli}}}{B}$, is an interaction at leading order in $N_{c}$, that arises due to Pauli repulsion at the quark level.  Note underlying point is that when two nucleons have an overlap, the Pauli principle ensures that the quark wave functions must distort from their free space value to ensure that single particle states are orthogonal.  It thus applies to the interactions of  neighboring baryons with the same spin-flavor structure.  For the toy problem, this is all pairwise interactions.   This term must be of $\mathcal{O}(N_{c})$, which should not be surprising; the Pauli repulsion fights against gluonic interactions between quarks whereby each baryon with $N_{c}$ constituent quarks interact via a color-Coulomb potential of $\mathcal{O}\left(\frac{1}{N_{c}}\right)$ with $N_{c}$ quarks of the neigboring baryons. The explicit form of the pairwise interaction  is as follows~\cite{3+1}:
\begin{equation}
\label{pauli}
\frac{E_{\textrm{Pauli}}}{B}=c_{1}N_{c}m_{q}\tilde{\alpha}_{s}^{2}\tilde{d}^{p}\exp\left(-c_{2}\tilde{d}\right) \left(1+\mathcal{O}\left(\frac{\ln\tilde{d}}{\tilde{d}}\right)\right) 
\end{equation}
providing the nucleons are asymptotically far from each other.  We will see {\it a posterori} that the saturation density density goes to zero at large $N_c$ and thus the particles are self-consistently in the regime of validity for Eq.~(\ref{pauli}).
In deriving the leading order contribution in $N_{c}$, a potential energy piece associated with interactions of neigboring baryons, which is of relative order $\mathcal{O}\left(\frac{\ln\tilde{d}}{\tilde{d}}\right)$ was ignored~\cite{3+1}. Here, $c_{1}=00245881$ and is a numerical factor that is proportional to the potential energy of an isolated baryon; $c_{2}=3.62275$ and $p=7.0107$; these factors are determined by the overlap of the neighboring wave functions and depend entirely on the tails of baryon wave functions~\cite{3+1}.  Note, moreover that the form of Eq.~(\ref{pauli}) is such that contribution from nearest neighbors will dominate.

The second piece, $E_{\textrm{gb}}$, is the interaction between two neigboring baryons at subleading order in $N_{c}$. This interaction happens via the exchange of scalar glueballs and can be well-approximated via a Yukawa potential. This description is exact only if the baryon charge densities are delta functions i.e. point sources in position space. Therefore, as long as saturation occurs for parametrically large distances, relative to the width of a baryon wave function, which is of $\mathcal{O}\left(\frac{1}{\tilde{\alpha}_{s}m_{q}}\right)$, using a Yukawa potential is justified. We will see \textit{a posteriori} that the point source assumption is indeed valid. The relative correction to the Yukawa potential is of $\mathcal{O}\left(\frac{1}{\tilde{d}}\right)$. Explicitly, the energy contribution to the glueball piece between neigboring baryons is:
\begin{equation}
\label{gb}
\frac{E_{\textrm{gb}}}{B}=-\tilde{g}_{\textrm{gb}}\frac{\exp\left(-\tilde{m}_{\textrm{gb}}\tilde{d}\right)}{\tilde{d}}\left(1+\mathcal{O}\left(\frac{1}{\tilde{d}}\right)\right )\ ,
\end{equation}
where $\tilde{m}_{\textrm{gb}}\equiv\frac{m_{\textrm{gb}}}{\tilde{\alpha}_{s}m_{q}}$ and $\tilde{g}_{\textrm{gb}}\equiv \tilde{\alpha}_{s}m_{q}g_{\textrm{gb}}$, with $g_{\textrm{gb}}$ being a dimensionless coupling constant; we further assume that it is positive definite, which guarantees that the glueball channel is attractive. Furthermore, note that the interaction is independent of $N_{c}$.

Here, both $\tilde{m}_{\textrm{gb}}$ and $\tilde{g}_{\textrm{gb}}$ are undetermined constants of $\mathcal{O}(N_{c}^{0})$. In principle, one can use lattice QCD at large values of $N_{c}$ to determine these constants. The mass of the lightest glueball can be determined by looking at the tail of the correlation function $\langle 0|F^{2}(x)F^{2}(0)|0\rangle$ for large $x$, with $F$ being the gluon field strength tensor. A determination of glueball masses in terms of the fundamental string constant has already been performed~\cite{teper}.  The glueball-baryon coupling, $g_{\textrm{gb}}$, is harder to determine. However, one could at least in principle extract interaction energies of two baryons (with heavy quarks) a fixed distance apart in the lattice; since we know the strength of the Pauli repulsion from Eq. (\ref{pauli}) and the lightest glueball mass from the lattice~\cite{teper}, we can determine $g_{\textrm{gb}}$.

Next we proceed to minimize the energy per baryon in the toy problem with respect to the dimensionless parameter, $\tilde{d}$, which gives us the following equation:
\begin{widetext}
\begin{equation}
%\left(c_{2}-\tilde{m}_{\textrm{gb}}\right)\tilde{d}-(p+1)\ln\tilde{d}+\ln\left(\tilde{m}_{\textrm{gb}}\tilde{d}+1\right)-\ln\left(c_{2}\tilde{d}-p\right)=\ln\left(\frac{N_{c}c_{1}m_{q}\tilde{\alpha}_{s}^{2}}{2N_{f}\tilde{g}_{\textrm{gb}}}\right)\ .
\left(c_{2}-\tilde{m}_{\textrm{gb}}\right)\tilde{d}-(p+1)\ln\tilde{d}+\ln\left(\tilde{m}_{\textrm{gb}}\tilde{d}+1\right)-\ln\left(c_{2}\tilde{d}-p\right)=\ln\left(\frac{N_{c}c_{1}m_{q}\tilde{\alpha}_{s}^{2}}{\tilde{g}_{\textrm{gb}}}\right)\ .
\label{min}
\end{equation}
\end{widetext}
Note that in constructing the energy expression for nuclear matter in the toy problem, we ignored corrections of relative $\mathcal{O}\left (\frac{\ln ( \tilde{d})}{\tilde{d}}\right)$. Therefore, in Eq~(\ref{min}), we proceed by ignoring terms of such order. We get:
\begin{widetext}
\begin{equation}
\label{d}
%\tilde{d}_{\textrm{SNM}}\left(1+\mathcal{O}\left(\frac{\ln\tilde{d}}{\tilde{d}}\right)\right)=\frac{1}{c_{2}-\tilde{m}_{\textrm{gb}}}\ln\left(\frac{N_{c}c_{1}c_{2}\tilde{\alpha}_{\textrm{s}}}{2 N_{f} g_{\textrm{gb}}\tilde{m}_{\textrm{gb}}}\right)\.
\tilde{d}_{\textrm{SNM}}^{\textrm{toy}}\left(1+\mathcal{O}\left(\frac{\ln\tilde{d}}{\tilde{d}}\right)\right)=\frac{1}{c_{2}-\tilde{m}_{\textrm{gb}}}\ln\left(\frac{N_{c}m_{q}}{\Lambda_{\textrm{QCD}}}\right)+
\ln\left(\frac{\Lambda_{\textrm{QCD}}c_{1}c_{2}\tilde{\alpha}_{s}^{2}}{g_{\textrm{gb}}m_{\textrm{gb}}}\right)\ .
\end{equation}
\end{widetext}
 $\tilde{\alpha}_{s}$ depends logarithmically on the quark mass, $c_{1}$ and $c_{2}$ are known to be of  $\mathcal{O}(N_{c}^{0}m_{q}^{0})$~\cite{3+1} and so is the scalar glueball mass, $m_{\textrm{gb}}$~\cite{teper}. Here, we have introduced $\Lambda_{\textrm{QCD}}$, which sets the scale for the quark masses. Furthermore, note that the separation between neighboring baryons at saturation, $\tilde{d}_{\textrm{SNM}}^{\textrm{toy}}$, is smaller for larger glueball masses. Additionally, $\tilde{d}_{\textrm{SNM}}^{\textrm{toy}}$, also decreases as the glueball-baryon coupling, $g_{\textrm{gb}}$, becomes larger.

Eq.~(\ref{d}) depends on an unknown but in principle knowable constant, $g_{\textrm{gb}}$. However, in the extreme large $N_{c}$ limit or the heavy quark mass limit where
\begin{equation}
\label{largeNclimit}
\left | \ln\left(\frac{N_{c}m_{q}}{\Lambda_{\textrm{QCD}}}\right)\right |\gg \left |\ln\left(\frac{\Lambda_{\textrm{QCD}}c_{1}c_{2}\tilde{\alpha}_{s}^{2}}{g_{\textrm{gb}}m_{\textrm{gb}}}\right)\right |\ ,
\end{equation}
and for quark masses heavy enough such that $c_{2}\gg\tilde{m}_{\textrm{gb}}$, separation between nucleons depends logarithmically on the number of colors and the quark mass relative to $\Lambda_{\textrm{QCD}}$:
\begin{equation}
\label{dSNM}
\tilde{d}_{\textrm{SNM}}^{\textrm{toy}}\approx\frac{1}{c_{2}}\ln\left(\frac{N_{c}m_{q}}{\Lambda_{\textrm{QCD}}}\right)\ ,\\
\end{equation}
with relative corrections, which are of the following orders:
\begin{equation}
\mathcal{O}\left( \frac{\ln\left(\frac{\Lambda_{\textrm{QCD}}c_{1}c_{2}\tilde{\alpha}_{s}^{2}}{g_{\textrm{gb}}m_{\textrm{gb}}}\right)}{\ln\left(\frac{N_{c}m_{q}}{\Lambda_{\textrm{QCD}}}\right))} \right)+\mathcal{O}\left( \frac{\tilde{m}_{\textrm{gb}}}{c_{2}}\right )\ .
\end{equation}

Having determined the separation of baryons at saturation, $\tilde{d}_{\textrm{SNM}}$, we proceed to determine the packing factor, $P$. Since the crystal structure will be fixed by the nearest neighbor distance, we can calculate the packing fraction by acting as though the baryons were hard spheres with a radius of half the nearest neighbor spacing. The  packing factor is defined as the fraction of volume that is occupied by the spheres associated with the baryons. The choice of packing factor is determined by which crystalline structure minimizes the energy per baryon. The structure that minimizes the energy per baryon must have the largest packing factor or the densest possible configuration. It has been known since Gauss (1831) that the largest possible packing factor for hard spheres has a value of $P_{\textrm{max}}=\frac{\pi}{\sqrt{18}}$, which is assumed by both a face-centered cubic and a hexagonal close packed structures. If the packing factor is lower than the maximum value $(P_{\textrm{max}})$, baryons on average will be further apart, which from Eqs~(\ref{en},\ref{pauli},\ref{gb}) means that the energy per baryon will be exponentially larger relative to the energy per baryon when packing factor assumes the maximum value. Therefore, the density of saturated nuclear matter in the toy problem, where we ignored spin-flavor degeneracy, follows straightforwardly; using the result of Eq~(\ref{dSNM}):
\begin{equation}\begin{split}
\label{densitySNM}
\rho_{\textrm{SNM}}^{\textrm{toy}} &=\frac{\sqrt{2}\tilde{\alpha}_{s}^{3}m_{q}^{3}}{\tilde{d}_{\textrm{SNM}}^{\textrm{toy}3}}\approx\sqrt{2}\left(\frac{c_{2}\tilde{\alpha}_{s}m_{q}}{{\ln \left( \frac{N_{c}m_{q}}{\Lambda_{\textrm{QCD}}}\right)}}\right)^{3}\ ,\\
{\textrm{where }} c_{2} & \approx 3.62275 \ .
\end{split}
\end{equation}
Here, $m_{q}$ is the quark mass and $\tilde{\alpha}_{s}$ is a coupling constant defined as $\tilde{\alpha}_{s}\equiv\frac{g^{2}N_{c}}{4\pi}$.
\\
\\
Let us  turn now to the real problem of nuclear matter in the double limit. In this case there is no restriction that nucleons are in a single spin-flavor configuration.
In the mean-field approximation, which is valid at large $N_{c}$~\cite{coherentstates}, all the $N_{c}$ quarks in a baryon must be in a single spin-flavor state.  There are thus $2 N_f$ possible states. In this general problem, baryons with the same spin-flavor configuration interact via Pauli repulsion and all baryons interact via scalar glueball exchange. Therefore, nuclear matter will form  a structure, whereby baryons with the same spin-flavor configuration form a crystal,  in which the repulsion due to the quark level Pauli effects and the attraction due to glueball exchange.   However, there are 2 $N_f$ of these crystal structures which can be superposed.  There is an energetic gain to superpose them on top of each other or nearly so.  This is because the Pauli repulsion does affect the baryons  with  different spin-flavor structures while the glueball-exchange attraction does.  The energy density of this system is  thus as follows:
\begin{equation}
\label{en2}
\frac{E_{\textrm{NM}}(\tilde{d})}{B}=\frac{E_{\textrm{Pauli}}(\tilde{d})}{B}+2N_{f}\frac{E_{\textrm{gb}}(\tilde{d})}{B}+\frac{E_{D}}{B}\ .
\end{equation}
The energy per baryon is different from that of the toy problem. Note that there are additional contributions to the energy per baryon. Firstly, $\frac{E_{D}}{B}$, is the energy (per baryon) due to the attractive  interaction between baryons of different spin-flavor configurations that sit directly on (or nearly) top of each other.  Note that this value cannot be easily calculated.  It is known to be of order unity but the precise value depends on details of short-distance interactions and not just the properties of the lightest glueball.  Indeed, we do not even know where exactly the minimum configuration has them on top of each other.  However, whatever configuration they ultimately have is independent of $d$ at large $d$. 
Note that there are glueball interactions between nearest neighbor baryons of different spin-flavor configurations; it is straightforward to see this glueball interaction energy per baryon is different from that of the toy problem by an additional factor of $2N_{f}$. Note that $\frac{E_{\textrm{Pauli}}}{B}$ and $\frac{E_{\textrm{gb}}}{B}$ are as defined in Eqs. ($\ref{pauli}$) and ($\ref{gb}$) respectively.

In order to determine the density of saturated nuclear matter for the full problem it is essential to understand the contribution to the energy per baryon through the term $\frac{E_{D}}{B}$. This term arises due to the interaction of baryon charges via  glueball exchanges. It does not depend on the separation of baryons, $d$, but only on the width of the baryon charge, which is of $\mathcal{O}(\frac{1}{\tilde{\alpha}_{s}m_{q}})$. In other words, the contribution due to $\frac{E_{\textrm{D}}}{B}$ is parametrically small compared to that of Pauli repulsion and glueball exchange between neighboring baryons. Therefore, the density of saturated nuclear matter for the full problem can be determined analogously to the toy problem modulo a change in the glueball-baryon coupling, $g_{\textrm{gb}}$:
\begin{equation}
\label{replacement}
g_{\textrm{gb}}\rightarrow2N_{f}g_{\textrm{gb}}\ .
\end{equation}
The above replacement takes into account the larger size of the Pauli repulsion (per baryon) in the full problem.

In the approximation defined in Eq.~(\ref{largeNclimit}) with the replacement of Eq.~(\ref{replacement}), the density of saturated nuclear matter is the same as that of Eq~(\ref{densitySNM}) i.e. the density in the toy problem, which is also independent of the glueball-baryon coupling. The density of saturated nuclear matter is as follows with the relevant corrections:
\begin{widetext}
\begin{equation}
\label{densitySNM}
\rho_{\textrm{SNM}} =\sqrt{2}\left(\frac{c_{2}\tilde{\alpha}_{s}m_{q}}{{\ln \left( \frac{N_{c}m_{q}}{\Lambda_{\textrm{QCD}}}\right)}}\right)^{3}\left(1+\mathcal{O}\left( \frac{\ln\left(\frac{\Lambda_{\textrm{QCD}}c_{1}c_{2}\tilde{\alpha}_{s}^{2}}{2N_{f}g_{\textrm{gb}}m_{\textrm{gb}}}\right)}{\ln\left(\frac{N_{c}m_{q}}{\Lambda_{\textrm{QCD}}}\right)} \right)+\mathcal{O}\left( \frac{\tilde{m}_{\textrm{gb}}}{c_{2}}\right )\right)\\
{\textrm{ with }} c_{2} \approx 3.62275.
\end{equation}
\end{widetext}
However, without doing substantially more work it is not possible to compute the energy per baryon of saturated nuclear matter; this is dominated by the contribution from $\frac{E_{\textrm{D}}}{B}$ which is unknown unless one can compute the relevant short-distance physics.

\begin{acknowledgements}
P.A. and T.D.C. acknowledge the support of the U.S. Department of Energy through grant number DEFG02-93ER-40762. I.D. acknowledges the hospitality and support of the TQHN research group at University of Maryland.
\end{acknowledgements}

\end{document}